\documentclass[prd,twocolumn,amsfonts,amssymb,nofootinbib]{revtex4}

\usepackage{graphicx}
\usepackage{subfig}

\usepackage{bm}

\newcommand{\beq}{\begin{equation}}
\newcommand{\eeq}{\end{equation}}
\newcommand{\beqs}{\begin{eqnarray}}\newcommand{\eeqs}{\end{eqnarray}}

\newcommand{\gsim}{\mathrel{\raisebox{-
.6ex}{$\stackrel{\textstyle>}{\sim}$}}}

\begin{document}

\title{Improved Lower Bounds on Partial Lifetimes for Nucleon Decay Modes}

\author{Sudhakantha Girmohanta and Robert Shrock}

\affiliation{ \ C. N. Yang Institute for Theoretical Physics and 
Department of Physics and Astronomy, \\
Stony Brook University, Stony Brook, NY 11794, USA }

\begin{abstract}

  In the framework of a baryon-number-violating effective Lagrangian, we
  calculate improved lower bounds on partial lifetimes for proton and bound
  neutron decays, including $p \to \ell^+ \ell'^+ \ell'^-$, $n \to \bar\nu
  \ell^+ \ell'^-$, $p \to \ell^+ \nu\bar\nu$, and $n \to \bar\nu \bar\nu \nu$,
  where $\ell$ and $\ell'$ denote $e$ or $\mu$, with both $\ell = \ell'$ and
  $\ell \ne \ell'$ cases.  Our lower bounds are substantially stronger than the
  corresponding lower bounds from direct experimental searches. We also present
  lower bounds on $(\tau/B)_{p \to \ell^+\gamma}$, $(\tau/B)_{n \to \bar\nu
    \gamma}$, $(\tau/B)_{p \to \ell^+ \gamma\gamma}$, and $(\tau/B)_{n \to
    \bar\nu \gamma\gamma}$. Our method relies on relating the rates for these
  decay modes to the rates for decay modes of the form $p \to \ell^+ M$ and $n
  \to \bar\nu M$, where $M$ is a pseudoscalar or vector meson, and then using
  the experimental lower bounds on the partial lifetimes for these latter
  decays.

\end{abstract}

\maketitle


\section{Introduction}
\label{intro_section}

Although the Standard Model (SM) conserves baryon number\footnote{ Recall that
  the violation of $B$ by SU(2)$_L$ instantons in the SM is negligibly small at
  temperatures low compared with the electroweak scale \cite{thooft}.}  $B$,
this is violated in many of its ultraviolet extensions. This violation is
natural in grand unified theories (GUTs) \cite{gg}-\cite{snowmass2013}, since
these theories place quarks and (anti)leptons in the same representation(s) of
the GUT gauge group.  More generally, baryon number violation (BNV) is expected
to occur in nature, because this is one of the necessary conditions for
explaining the observed baryon asymmetry of the universe \cite{sakharov}.  A
number of dedicated experiments have been carried out since the early 1980s to
search for proton decay and the decay of neutrons bound in nuclei. These
experiments have obtained null results and have set resultant stringent upper
limits for the rates of nucleon decays\footnote{We shall use the term ``nucleon
  decay'' to mean the decay of a proton or the $B$-violating decay of a neutron
  bound in a nucleus.}.

In this paper, within the framework of a baryon-number-violating effective
Lagrangian, ${\cal L}_{eff}$, we shall calculate improved lower bounds on
partial lifetimes for a number of nucleon decays.  Let us denote the rate for
the decay of a nucleon $N$ (where $N=p$ or $n$) to a final state $f.s.$ as
$\Gamma_{N \to f.s.}$, which is the inverse of the partial lifetime,
$(\tau/B)_{N \to f.s.}  = [\Gamma_{N \to f.s.}]^{-1}$, where $B$ denotes the
branching ratio for this decay mode. Our method is to derive an approximate
relation between the rate $\Gamma_{N \to f.s.}$ for the decay of a nucleon to a
final state $f.s.$, $N \to f.s.$ (where $N=p$ or $n$) and the rate for the
decay $\Gamma_{N \to f.s.'}$ to a different final state, denoted $f.s.'$.
Combining this relation with the experimental lower bound on $(\tau/B)_{N \to
  f.s.}$, we derive an approximate lower bound on $(\tau/B)_{N \to f.s.'}$ for
each final state $f.s.'$.  Our theoretical framework is minimal, in the sense
that the only physics beyond the SM (BSM) is that is assumed is that which
produces the basic set of local four-fermion operators in ${\cal L}_{eff}$. 
Although the lower bounds that we derive are only approximate, they
are useful because for many final states $f.s.'$ they are more stringent than
the lower bounds on $(\tau/B)_{N \to f.s.'}$ from direct experimental searches.

This paper is organized as follows.  In
Sect. \ref{effective_lagrangian_section} we discuss the four-fermion operators
in ${\cal L}_{eff}$. We present our lower bounds on $(\tau/B)$ for $p \to
\ell^+ \gamma$ and $n \to \bar\nu \gamma$ in Sect.  \ref{p_to_lgamma_section}.
Sects. \ref{p_to_3l_section} and \ref{n_to_nu2l_section} contain our lower
bounds on $(\tau/B)$ for $p \to \ell^+ \ell'^+ \ell'^-$ and $n \to \bar\nu
\ell^+ \ell^-$, where $\ell$ and $\ell'$ denote $e$ or $\mu$, including both
$\ell = \ell'$ and $\ell \ne \ell'$ cases.  In Sects. \ref{p_to_l2nu_section}
and \ref{n_to_3nu_section} we derive lower bound bounds on $(\tau/B)$ for $p
\to \ell^+ \nu \bar\nu$ and $n \to \bar\nu \bar\nu \nu$. Here and below we use
a symbolic notation in which $\nu$ may refer to an electroweak-doublet neutrino
of some generation, or to an electroweak-singlet neutrino; the context will
make clear the meaning. In Sect. \ref{other_decays_section} we remark on the
application of our method to other decays, including $p \to \ell^+
\gamma\gamma$ and $n \to \bar\nu \gamma\gamma$. Our conclusions are presented
in Sect.  \ref{conclusion_section} and some relevant phase-space formulas are
given in Appendix \ref{phase_space_appendix}.


\section{Effective Lagrangian}
\label{effective_lagrangian_section}

Given the established experimental upper bounds on the rates for nucleon
decays, it follows that the mass scale(s) characterizing the
baryon-number-violating physics responsible for these days must be larger than
the electroweak-symmetry-breaking (EWSB) scale, $v \simeq 250$ GeV.  Hence, one
can analyze these decays using an effective Lagrangian, ${\cal L}_{eff}$, that
is invariant not only with respect to color SU(3)$_c$, but also with respect 
to gauge transformations of the electroweak gauge group, 
$G_{EW} = {\rm SU}(2)_L \otimes {\rm U}_Y$.  With
the original SM fermions, before the addition of any electroweak-singlet
neutrinos, the four-fermion operators ${\cal O}_i$ in ${\cal L}_{eff}$ that
contribute to nucleon decays are as follows. We denote $Q^\alpha_{a,L} = {Q^{1
    \alpha}_a \choose Q^{2 \alpha}_a}_L = {u_a^\alpha \choose d_a^\alpha}_L$
and $L_{a,L} = {L^1_a \choose L^2_a }_L = {\nu_{\ell_a} \choose \ell_a}_L$,
where $\alpha, \ \beta, \ \gamma$ are SU(3)$_c$ indices, 
and $a$ is a generation index, with $d^\alpha_1 = d^\alpha$, 
$d^\alpha_2 = s^\alpha$, $d^\alpha_3=b^\alpha$, 
$\ell_1=e$, $\ell_2=\mu$, $\ell_3=\tau$, etc.  The operators
contributing to proton decay are\footnote{Two other operators vanish
  identically in the case $a_1=a_2=a_3=1$ relevant for nucleon decay, namely
  ${\cal O}_5 = \epsilon_{\alpha\beta\gamma} [u^{\alpha \ T}_{a_1,R} C
  u^{\beta}_{a_2,R}] [d^{\gamma \ T}_{a_3,R} C \ell_{a_4,R}]$ and ${\cal O}_6 =
  (\epsilon_{ik}\epsilon_{jm} + \epsilon_{im}\epsilon_{jk})
  \epsilon_{\alpha\beta\gamma} [Q^{i \alpha \ T}_{a_1,L} C Q^{j \beta}_{a_2,L}]
  [Q^{k \gamma \ T}_{a_3,L} C L^m_{a_4,L}]$.} \cite{weinberg79,wz79}
\beq
{\cal O}_1 = \epsilon_{\alpha\beta\gamma}
[u^{\alpha \ T}_{a_1,R} C d^\beta_{a_2,R}]
[u^{\gamma \ T}_{a_3,R} C \ell_{a_4,R}]
\label{op1}
\eeq
\beq
{\cal O}_2 =  \epsilon_{ij} \epsilon_{\alpha\beta\gamma}
[Q^{i \alpha \ T}_{a_1,L} C Q^{j \beta}_{a_2,L}]
[u^{\gamma \ T}_{a_3,R} C \ell_{a_4,R}]
\label{op2}
\eeq
\beq
{\cal O}_3 = \epsilon_{km}\epsilon_{\alpha\beta\gamma}
[u^{\alpha \ T}_{a_1,R} C d^\beta_{a_2,R}]
[Q^{k \gamma \ T}_{a_3,L} C L^m_{a_4,L}]
\label{op3}
\eeq
and
\beq
{\cal O}_4 = \epsilon_{ij} \epsilon_{km} \epsilon_{\alpha\beta\gamma}
[Q^{i \alpha \ T}_{a_1,L} C Q^{j \beta}_{a_2,L}]
[Q^{k \gamma \ T}_{a_3,L} C L^m_{a_4,L}] \ , 
\label{op4}
\eeq
where $C$ is the Dirac charge conjugation matrix satisfying $C \gamma_\mu
C^{-1} = -(\gamma_\mu)^T$, $C=-C^T$ and $i,\ j, \ k, \ m$ are SU(2)$_L$
indices. As noted in \cite{weinberg79}, four-fermion operators with bilinears
involving Dirac vector and tensor operators $\gamma_\mu$ and $\sigma_{\mu\nu} =
(i/2)[\gamma_\mu,\gamma_\nu]$ can be transformed to the operators listed above
via Fierz identities.  These operators have $\Delta B=-1$ and $\Delta L=-1$,
where $L$ denotes total lepton number.

After the discovery of nonzero neutrino masses and lepton flavor mixing, a
natural generalization of the Standard Model has involved the introduction of a
set of electroweak-singlet neutrinos $\nu_{s,R}$, $s=1,...,n_s$, which are
necessary to form Dirac neutrino mass terms via Yukawa couplings $\sum_{a=1}^3
\sum_{s=1}^{n_s} y_{as} [\bar L_{a,L} \nu_{s,R} \tilde \phi] + h.c.$, where
$\tilde \phi \equiv i\sigma_2 \phi^*$ and $\phi = {\phi^+ \choose \phi^0}$ is
the SM Higgs doublet. These $\nu_{s,R}$ neutrinos also generically form
Majorana bare mass terms $\sum_{s,s'=1}^{n_s}M^{(\nu)}_{s,s'} \nu_{s,R}^T C
\nu_{s',R} + h.c.$, thereby explicitly breaking total lepton number $L$ by 2
units. With the inclusion of these $\nu_{s,R}$, there are two additional types
of operators for nucleon decay, namely
\beq
{\cal O}_7 = \epsilon_{\alpha\beta\gamma}
[u^{\alpha \ T}_{a_1,R} C d^\beta_{a_2,R}]
[d^{\gamma \ T}_{a_3,R} C \nu_{s,R}]
\label{op7}
\eeq
and
\beqs
{\cal O}_8 &=& \epsilon_{ij}\epsilon_{\alpha\beta\gamma}
[Q^{i \alpha \ T}_{a_1,L} C Q^{j \beta}_{a_2,L}]
[d^{\gamma \ T}_{a_3,R} C \nu_{s,R}] \ . 
\label{op8}
\eeqs
The generation indices $(a_1,a_2,a_3,a_4)$ in 
the ${\cal O}_r$ with $1 \le r \le 4$ and the 
indices $(a_1,a_2,a_3,s)$ in ${\cal O}_r$ with $r=7, 8$ will be left 
implicit in the notation. 

In terms of these fields, a minimal low-energy effective 
Lagrangian giving rise to nucleon decay can be written as 
\beq
{\cal L}_{eff} = \sum_{r} \sum_{\{ a_i \} ;s} c_r {\cal O}_r   \ , 
\label{leff}
\eeq
where the second sum is over all of the generation indices $(a_1,a_2,a_3,a_4)$
in the operators ${\cal O}_r$, $1 \le r \le 4$ and the indices
$(a_1,a_2,a_3,s)$ in ${\cal O}_7$ and ${\cal O}_8$.  Since these operators
${\cal O}_r$ have Maxwellian dimension 6 (in mass units), the coefficients
$c_r$ have dimension $-2$, and we write $c_r = \bar c_r/(M_{BNV})^2$, where
$M_{BNV}$ denotes an effective mass scale characterizing the baryon-number
violation.  In general, $\bar c_r$ depends on the generational indices of
fermion fields in ${\cal O}_r$; this is again left implicit in the notation.


\section{The Decays $p \to \ell^+ \gamma$ and $n \to \bar\nu \gamma$}
\label{p_to_lgamma_section}

\begin{figure*}
  \begin{center}
    \subfloat[]{
       \includegraphics[width=0.3\textwidth]{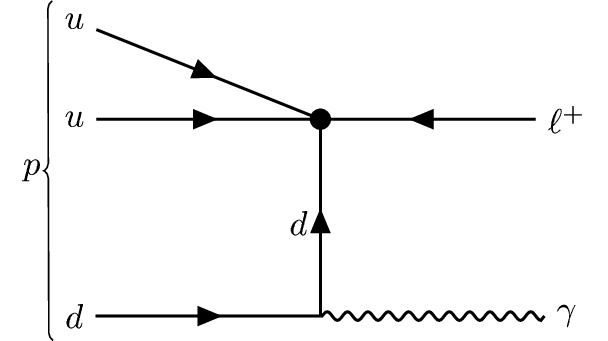}
     }
     \subfloat[]{
       \includegraphics[width=0.3\textwidth]{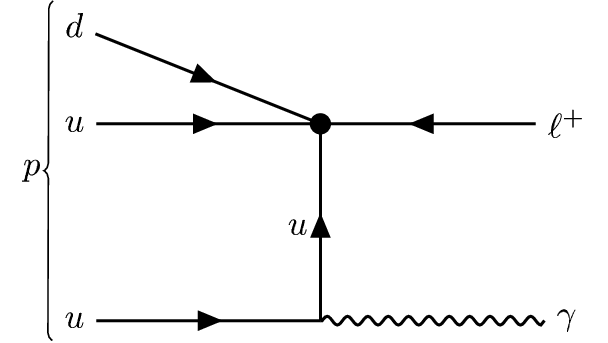}
     }
       \subfloat[]{
       \includegraphics[width=0.3\textwidth]{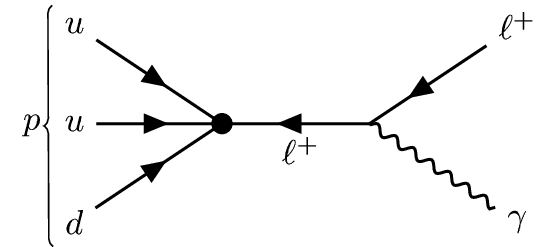}
       }
   \end{center}
\caption{Feynman diagrams for $p \to \ell^+ \gamma$ with
$\ell^+ = e^+, \ \mu^+$.}
\label{p_to_l_gamma_figure}
\end{figure*}

We begin by deriving approximate lower limits, within this theoretical
framework, on the partial lifetimes for the decays $p \to \ell^+ \gamma$ and $n
\to \bar\nu\gamma$, where $\ell^+ = e^+$ or $\mu^+$ and $\bar\nu$ may be an
electroweak-nonsinglet antineutrino, $\bar\nu_e$, $\bar\nu_\mu$, or 
$\bar\nu_\tau$, or an electroweak-singlet antineutrino, $\bar\nu_s$ where $1
\le s \le n_s$.\footnote{In general, the EW-singlet interaction eigenstate
  $\bar\nu_s$ is a linear combination of mass eigenstates. Here the statement
  refers to the mass eigenstates with masses that are sufficiently small so
  that they are kinematically allowed to occur in the decay. This is also
  understood for the EW-nonsinglet $\bar\nu_\ell$.}  In view of these
possibilities, we omit a subscript on $\bar\nu$ here and in similar cases
below. With our ${\cal L}_{eff}$, the leading contributions to the decay $p \to
\ell^+ \gamma$ arise from the diagrams in Fig. \ref{p_to_l_gamma_figure}.
Fig. \ref{p_to_l_gamma_figure}(a) shows a process in which $\sum_r c_r {\cal
  O}_r$ in ${\cal L}_{eff}$ 
(represented by the blob at the four-fermion vertex) with $1 \le r
\le 4$ and $a_1=a_2=a_3=1$, $a_4=1$ or 2 for $\ell^+=e^+$ or $\mu^+$,
transforms an initial $uu$ pair in the proton to $\ell^+ d^c$, and the $d^c$
annihilates with the $d$ in the proton to produce the outgoing photon. This
figure is also understood to include a process in which the $d$ quark in the
proton emits the photon, transitioning to a virtual $d$ that undergoes the BNV
process depicted by the blob, resulting in the outgoing $\ell^+$.
Fig. \ref{p_to_l_gamma_figure}(b) shows the analogous processes involving the
BNV transformation $du \to \ell^+ u^c$.  In Fig. \ref{p_to_l_gamma_figure}(c),
the initial $uud$ quarks in the proton are transformed via the BNV ${\cal
  L}_{eff}$ to a virtual $s$-channel $\ell^+$ that then radiates the photon.
%
%
%
%

\begin{figure*}
  \begin{center}
  \subfloat[]{
       \includegraphics[width=0.3\textwidth]{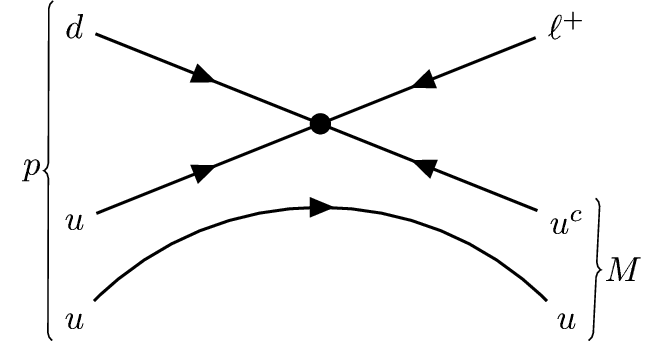}
     }
     \hspace{2 cm}
   \subfloat[]{
       \includegraphics[width=0.3\textwidth]{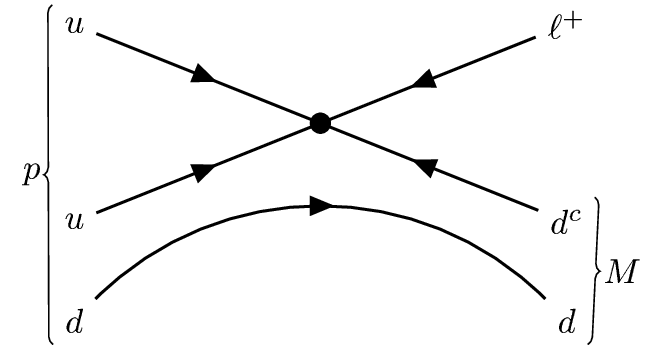}
     }
  \end{center}
\caption{Feynman diagrams for $p \to \ell^+ M$, where
$\ell^+=e^+, \ \mu^+$ and $M$ denotes a pseudoscalar or vector meson.}
\label{p_to_l_meson_figure}
\end{figure*}

To proceed, we relate the amplitude for the $p \to \ell^+ \gamma$ decay to an
amplitude for $p \to \ell^+ M$ decay, where $\ell^+ = e^+$ or $\mu^+$ and 
$M$ denotes a neutral meson containing light quarks, such as 
$\pi^0$, $\eta$, $\rho^0$, or $\omega$.  Feynman 
diagrams for the decay $p \to \ell^+ M$ are shown in Fig. 
\ref{p_to_l_meson_figure}. 
%
%
%
Both the isoscalar mesons $\eta$ and $\omega$ and the isovector mesons $\pi^0$
and $\rho^0$ are relevant for this relation, since the electromagnetic current
has both isoscalar and isovector parts, as embodied in the relation $Q_{em} =
I_3 + (Y/2)$, where here $I$ and $Y$ denote isospin and hypercharge. The final
states involving the vector mesons $\omega$ and $\rho^0$ share with the final
state involving the photon the property that they are all vector $(J=1$)
particles.  On the other hand, the final states involving the $\pi^0$ and
$\eta$ have (mass)$^2$ values 0.0182 GeV$^2$ and 0.301 GeV$^2$ that are smaller
than the (mass)$^2$ values of 0.602 GeV$^2$ and 0.613 GeV$^2$ for the $\omega$
and $\rho^0$ and hence closer to the zero mass of the photon.  In view of the
complementary similarities (in spin and (mass)$^2$) of the $\omega$ and
$\rho^0$ hadronic final states, on the one hand, and the $\pi^0$ and $\eta$
hadronic final states, on the other, to the photon, we shall use all of these
decay modes for our comparison.  We list the experimental lower bounds on the
partial lifetimes $(\tau/B)$ for relevant proton and bound neutron decay
modes in Tables \ref{proton_decay_table} and \ref{neutron_decay_table}. These
and other limits listed here are at the 90 \% confidence level. 
\begin{table}
  \caption{\footnotesize{List of (a) experimental lower bounds (at the 90 \%
      CL) on $(\tau/B)$, denoted $(\tau/B)_{p \to \ell^+ M; 
      {\rm exp. \ l.bnd.}}$, for various proton decays of the form
      $p \to \ell^+ M$, where $\ell^+ =e^+$ or $\mu^+$ and $M=\pi^0, \ 
      \eta, \ \rho^0, \ \omega$, with references, given in the first to third 
      columns; (b) in the fourth column, our resultant 
      estimated lower bounds on 
      $(\tau/B)$ for $p \to \ell^+ \gamma$ obtained from Eq.
      (\ref{tau_p_to_lgamma_wrt_lmeson}), denoted 
      $(\tau/B)_{p \to \ell^+ \gamma; {\rm est. \ l.bnd.}}$. 
      All limits on $(\tau/B)$ are given in units of $10^{33}$ yr. See text
      for further discussion.}}
\begin{center}
\begin{tabular}{|c|c|c|c|} \hline\hline
$p \to \ell^+ M$ & $(\tau/B)_{p \to \ell^+ M; {\rm exp. \ l.bnd.}}$ & 
Ref. &  $(\tau/B)_{p \to \ell^+ \gamma; {\rm est. \ l.bnd.}}$ \\
\hline 
$p \to e^+ \pi^0$     & 16.0   & \cite{abe17}     & $2 \times 10^2$    \\
$p \to \mu^+ \pi^0$   & 7.7    & \cite{abe17}     & $0.8 \times 10^2$  \\
$p \to e^+ \eta$      & 10.0   & \cite{abe17d}    & $0.7 \times 10^2$  \\
$p \to \mu^+ \eta$    & 4.7    & \cite{abe17d}    & 30   \\
$p \to e^+ \rho^0$    & 0.720  & \cite{abe17d}    & 2    \\
$p \to \mu^+ \rho^0$  & 0.570  & \cite{abe17d}    & 1  \\ 
$p \to e^+ \omega$    & 1.60   & \cite{abe17d}    & 5    \\
$p \to \mu^+ \omega$  & 2.80   & \cite{abe17d}    & 7    \\
\hline\hline
\end{tabular}
\end{center}
\label{proton_decay_table}
\end{table}


\begin{table}
  \caption{\footnotesize{List of (a) experimental lower bounds (at the 90 \%
      CL) on $(\tau/B)$, denoted $(\tau/B)_{n \to \bar\nu M; 
      {\rm exp. \ l.bnd.}}$, 
      for various baryon-number-violating neutron decays of the form
      $n \to \bar\nu M$, where $M=\pi^0, \ \eta, \ \rho^0, \ \omega$, with
      references, given in the first to third columns; 
      (b) in the fourth column, our resultant estimated lower bounds on 
      $(\tau/B)$ for $n \to \bar\nu \gamma$ obtained from Eq.
      (\ref{tau_n_to_nubar_gamma_wrt_nubar_meson}), denoted 
      $(\tau/B)_{n \to \bar\nu \gamma; {\rm est. \ l.bnd.}}$. 
      All limits on $(\tau/B)$ are given in units of $10^{33}$ yr. See text
      for further discussion.}}
\begin{center}
\begin{tabular}{|c|c|c|c|} \hline\hline
$n \to \bar\nu M$ & $(\tau/B)_{n \to \bar\nu M; {\rm exp. \ l.bnd.}}$ & 
Ref. &  $(\tau/B)_{n \to \bar\nu \gamma; {\rm est. \ l.bnd..}}$ \\
\hline 
$n \to\bar\nu \pi^0$  & 1.1          & \cite{abe14e}     & 10  \\
$n \to\bar\nu \eta $  & 0.158        & \cite{mcgrew99}   & 1  \\
$n \to\bar\nu \omega$ & 0.108        & \cite{mcgrew99}   & 0.4   \\
$n \to\bar\nu \rho^0$ & $1.9 \times 10^{-2}$ & \cite{seidel88}  &  0.07 \\
\hline\hline
\end{tabular}
\end{center}
\label{neutron_decay_table}
\end{table}
%
The phase space factor for a decay of a nucleon $N$ to a two-body final state
with particles of masses $m_1$ and $m_2$ is given by Eq. (\ref{r2}) in Appendix
\ref{phase_space_appendix}. We list the values of 
$(8\pi)R^{(f.s.)}_2 = [\lambda(1,(m_1/m_N)^2,(m_2/m_N)^2]^{1/2}$ in Table
\ref{2body_phase_space_table} for nucleon decays to various final states $f.s.$
of relevance here.  
\begin{table}
  \caption{\footnotesize{Reduced two-body phase-space factors 
 $(8\pi)R_2^{(f.s.)}$ 
      for two-body proton decays to the indicated final states ($f.s.$).}}
\begin{center}
\begin{tabular}{|c|c|} \hline\hline
decay &  $(8\pi)R_2^{(f.s.)}$ \\
\hline 
$p \to e^+ \gamma$    &  1.000  \\
$p \to \mu^+ \gamma$  &  0.987  \\
$p \to e^+ \pi^0$     &  0.979  \\
$p \to \mu^+ \pi^0$   &  0.966  \\
$p \to e^+ \eta $     &  0.658  \\
$p \to \mu^+ \eta$    &  0.632  \\
$p \to e^+ \rho^0$    &  0.316  \\
$p \to \mu^+ \rho^0$  &  0.241  \\ 
$p \to e^+ \omega$    &  0.304  \\
$p \to \mu^+ \omega$  &  0.222  \\
\hline
$n \to\bar\nu \gamma$ &  1.000  \\
$n \to\bar\nu \pi^0$  &  0.979  \\
$n \to\bar\nu \eta$   &  0.659  \\
$n \to\bar\nu \rho^0$ &  0.318  \\
$n \to\bar\nu \omega$ &  0.306  \\
\hline\hline
\end{tabular}
\end{center}
\label{2body_phase_space_table}
\end{table}
%

Because the hadronic matrix elements $\langle M | {\cal L}_{eff} | p\rangle$
and $\langle 0 | {\cal L}_{eff} | p\rangle$ that enter in the respective $p \to
\ell^+ M$ and $p \to \ell^+ \gamma$ decays are different and the coefficients
$c_r$ that enter into ${\cal L}_{eff}$ depend on the UV completion of the 
Standard Model that is responsible for the baryon-number violation, we will 
restrict our analysis to a rough estimate of the relation between the
corresponding decays\footnote{For lattice calculations of hadronic matrix
elements, see Ref. \cite{aoki}.}  A similar comment applies to $n \to \bar\nu
\gamma$ decays. We have 
\beq
\Gamma_{p \to \ell^+ \gamma} \sim e^2 \, 
\bigg [ \frac{R^{(\ell^+ \gamma)}_2}{R^{(\ell^+ M)}_2} \bigg ] 
\, \Gamma_{p \to \ell^+ M} 
\ , 
\label{p_to_lgamma_wrt_lmeson}
\eeq
where $\ell^+ = e^+$ or $\mu^+$ and $e^2=4\pi \alpha_{em}$ is the squared
electromagnetic coupling.  The ratio of phase-space factors is included to take
account of the difference in phase space for the $p \to \ell^+ \gamma$ and $p
\to \ell^+ M$ decay modes.  Equivalently,
\beq
(\tau/B)_{p \to \ell^+ \gamma} \sim (4\pi \alpha_{em})^{-1} \, 
\bigg [ \frac{R^{(\ell^+ M)}_2}{R^{(\ell^+ \gamma)}_2} \bigg ] 
\, (\tau/B)_{p \to \ell^+ M}\ . 
\label{tau_p_to_lgamma_wrt_lmeson}
\eeq
Note that although the branching ratios for the various decay modes in Eq. 
(\ref{tau_p_to_lgamma_wrt_lmeson}) depend on the UV completion of the SM, 
the basic relation (\ref{p_to_lgamma_wrt_lmeson}) is between the absolute rates
themselves, which does not depend on these branching ratios. 

We next make use of the experimental lower bounds on the partial lifetimes for
various relevant proton decay modes, as displayed in Table
\ref{proton_decay_table}. These lower bounds on $(\tau/B)$ are all from the
SuperKamiokande (SK) experiment; the bounds for $p \to \ell^+ \pi^0$ and $p \to
\ell^+ \eta$ are from Ref. \cite{abe17}, while the bounds for $p \to \ell^+
\omega$ and $p \to \ell^+ \rho$ are from Ref. \cite{abe17d}.  In the right-most
column of Table \ref{proton_decay_table}, we list our estimates for the lower
bounds on $p \to \ell^+ \gamma$ obtained by combining the relation
(\ref{tau_p_to_lgamma_wrt_lmeson}) with the experimental lower bounds on
$(\tau/B)_{p \to \ell^+ M}$ given in the middle column, for $\ell^+=e^+$ and
$\ell^+ = \mu^+$. In view of the approximate nature of our estimated lower
bounds on $(\tau/B)_{p \to \ell^+ \gamma}$ in Table \ref{proton_decay_table},
we list these only to one significant figure, and we follow this format with
our estimates for other nucleon decay modes below.

Our estimates for lower bounds on $(\tau/B)_{p \to \ell^+ \gamma}$ may be
compared with lower bounds from direct experimental searches, which are as 
follows. The IMB-3 experiment obtained the limits \cite{mcgrew99}
$(\tau/B)_{p \to e^+ \gamma} > 0.670 \times 10^{33}$ yr. 
and $(\tau/B)_{p \to \mu^+ \gamma} > 0.478 \times 10^{33}$ yr. 
More recently, the SK experiment has reported the limits \cite{sussman18} 
$(\tau/B)_{p \to e^+ \gamma} > 4.1 \times 10^{34}$ yr. and
$(\tau/B)_{p \to \mu^+ \gamma} > 2.4 \times 10^{34}$ yr. 
These comparisons are summarized in Table
\ref{nucleon_decay_table}. In this table, for each of the decay modes 
$p \to \ell^+ \gamma$ and $n \to \bar\nu \gamma$, we list the range of
estimated lower that we obtain using Eq. (\ref{tau_p_to_lgamma_wrt_lmeson})
with all of the input bounds for $p \to \ell^+ M$. 
%
\begin{table}
  \caption{\footnotesize{Table listing (a) experimental lower bounds (l.bnd.)
  (at the 90 \% CL) on $(\tau/B)$ for various nucleon decays, denoted 
  $(\tau/B)_{\rm exp. \ l.bnd.}$ with references, given in
  the first to third columns; and 
  (b) our theoretical estimated lower bounds on the partial lifetimes
  for these nucleon decays, denoted  $(\tau/B)_{\rm est. \ l.bnd.}$. 
  The units of $(\tau/B)$ are $10^{33}$ yr. The abbreviation NA means ``not
  available''.  See text for further details.}}
\begin{center}
\begin{tabular}{|c|c|c|c|} \hline\hline
decay mode & $(\tau/B)_{\rm exp. \ l.bnd.}$ & Ref. &  
             $(\tau/B)_{p \to \ell^+ \gamma; {\rm est. \ l.bnd..}}$ \\
\hline 
$p \to e^+ \gamma$          & 41     & \cite{sussman18}  & $\sim 10-10^2$  \\
$p \to \mu^+ \gamma$        & 24     & \cite{sussman18}  & $\sim 10-10^2$  \\ 
$p \to e^+ \gamma\gamma$    & 1.00   & \cite{berger91}   & $\sim 10^4$     \\
$p \to \mu^+ \gamma\gamma$  & NA     & NA                & $\sim 10^4$     \\
$n \to \bar\nu \gamma$      & 0.55   & \cite{takhistov15} & $\sim 1-10$  \\
$n\to\bar\nu \gamma\gamma$   & 2.19& \cite{mcgrew99}  & $\sim 10^3$ \\
$p \to e^+ e^+e^-$          & 0.793  & \cite{mcgrew99}  & $\sim 10^4$  \\
$p \to \mu^+ e^+e^-$        & 0.529  & \cite{mcgrew99}  & $\sim 10^4$  \\
$p \to e^+ \mu^+\mu^-$      & 0.359  & \cite{mcgrew99}  & $\sim 10^4$  \\
$p \to \mu^+ \mu^+\mu^-$    & 0.675  & \cite{mcgrew99}  & $\sim 10^4$  \\
$n \to \bar\nu e^+ e^-$     & 0.257  & \cite{mcgrew99}  & $\sim 10^3$   \\
$n \to \bar\nu \mu^+\mu^-$  & 0.079  & \cite{mcgrew99}  & $\sim 10^3$   \\
$n \to \bar\nu \mu^+ e^-$   & 0.083  & \cite{mcgrew99}  & $\sim 10^{11}$  \\
$p \to \bar\nu e^+ \nu_e$   & 0.17   & \cite{takhistov14} & $\sim 10^{12}$   \\
$p \to\bar\nu\mu^+\nu_\mu$  & 0.22   & \cite{takhistov14} & $\sim 10^{12}$ \\
$n \to \bar\nu \nu \bar\nu$ & $0.58 \times 10^{-3}$&\cite{araki06} & 
$\sim 10^{11}$ \\
\hline\hline
\end{tabular}
\end{center}
\label{nucleon_decay_table}
\end{table}
%

\begin{figure*}
  \begin{center}
 \subfloat[]{
       \includegraphics[width=0.3\textwidth]{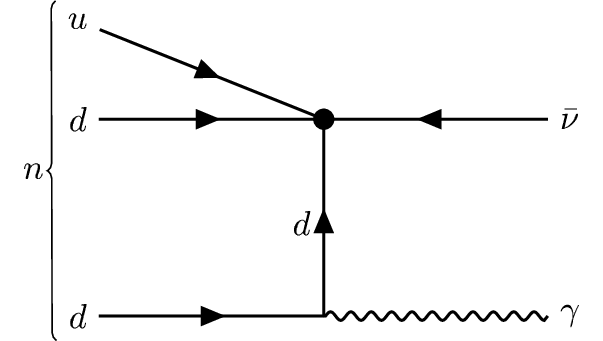}
     }
     \hspace{1.5cm}
    \subfloat[]{
       \includegraphics[width=0.3\textwidth]{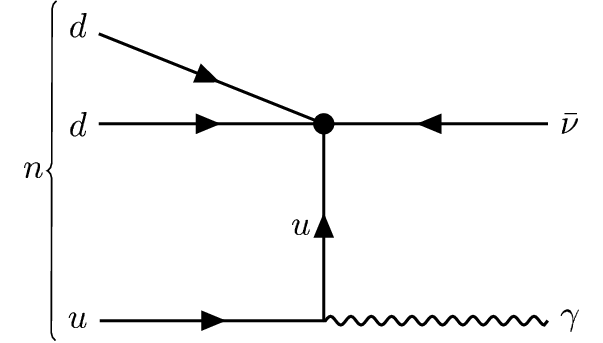}
     }
  \end{center}
\caption{Feynman diagrams for $n \to \bar\nu \gamma$.}
\label{n_to_nubar_gamma_figure}
\end{figure*}

\begin{figure*}
  \begin{center}
 \subfloat[]{
       \includegraphics[width=0.3\textwidth]{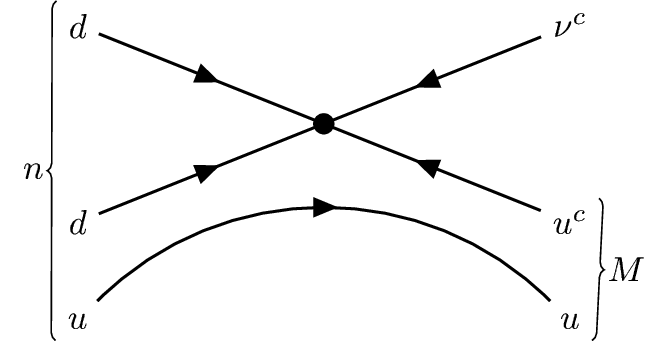}
     }
     \hspace{2 cm}
   \subfloat[]{
       \includegraphics[width=0.3\textwidth]{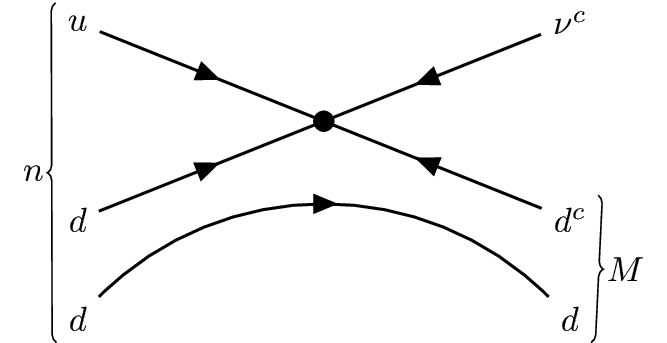}
     }
  \end{center}
\caption{Feynman diagrams for $n \to \bar\nu M$, where
$M$ denotes a pseudoscalar or vector meson.}
\label{n_to_nubar_meson_figure}
\end{figure*}

We proceed to carry out the corresponding analysis for the decay $n \to \bar\nu
\gamma$. The leading Feynman diagrams contributing to this decay are shown in
Fig. \ref{n_to_nubar_gamma_figure}. 
%
%
%
%
As noted above, since the antineutrino is not observed, it could either be an
electroweak-nonsinglet, $\bar\nu_\ell$ (i.e., $\bar\nu_e$, $\bar\nu_\mu$, or
$\bar\nu_\tau$) or an EW-singlet, $\bar\nu_s$.  The leading contributions to
the $n \to \bar\nu_\ell \gamma$ decay arise from the relevant terms in ${\cal
  O}_3$ and ${\cal O}_4$ with $a_1=a_2=a_3=1$ and arbitrary $a_4$ (where
$a_4=1, \ 2, \ 3$ for $\bar\nu_e$, $\bar\nu_\mu$, and $\bar\nu_\tau$), namely
the term $-\epsilon_{\alpha\beta\gamma} [u^{\alpha \ T}_R C
d^\beta_R][d^{\gamma \ T}_L C \nu_{a_4,L}]$ in ${\cal O}_3$ and the term
$-2\epsilon_{\alpha\beta\gamma} [u^{\alpha \ T}_L C d^\beta_L][d^{\gamma \ T}_L
C \nu_{a_4,L}]$ in ${\cal O}_4$.  The leading contributions to $n \to \bar\nu_s
\gamma$ arise from the term $\epsilon_{\alpha\beta\gamma} [u^{\alpha \ T}_R C
d^\beta_R][d^{\gamma \ T}_R C \nu_{s,R}]$ in ${\cal O}_7$ and the term
$2[u^{\alpha \ T}_L C d^\beta_L][d^{\gamma \ T}_R C \nu_{s,R}]$ in ${\cal
  O}_8$.  Our procedure is again to obtain an approximate relation between the
rates for $n \to \bar\nu \gamma$ and for $n \to \bar\nu M$.  In
Figs. \ref{n_to_nubar_meson_figure}(a) and \ref{n_to_nubar_meson_figure}(b), we
show diagrams contributing to the decay $n \to \bar\nu M$. By the same method
as we used above, we obtain the approximate estimate
\beqs
(\tau/B)_{n \to \bar\nu \gamma} & \sim & (4\pi \alpha_{em})^{-1} \, 
\bigg [ \frac{R^{(\bar\nu M)}_2}{R^{(\bar\nu \gamma)}_2} \bigg ] \, 
(\tau/B)_{n \to \bar\nu M}\  \cr\cr
& \sim & (4\pi \alpha_{em})^{-1} \, \Big [1-\frac{m_{M}^2}{m_n^2} \Big ] \, 
(\tau/B)_{n \to \bar\nu M} \ . \cr\cr
&&
\label{tau_n_to_nubar_gamma_wrt_nubar_meson}
\eeqs
We make use of the experimental lower bounds on $(\tau/B)$ for relevant decays
$n \to \bar\nu M$, displayed in Table \ref{neutron_decay_table}.\footnote{Since
  the experiments do not observe the $\bar\nu$, these experimental bounds are
  more general. For example, the lower bound on the partial lifetime for $n \to
  \bar\nu \pi^0$ actually applies to any decay of the form $n \to x^0 \pi^0$,
  where $x^0$ is a neutral, weakly interacting particle or antiparticle that
  does not decay in the detector, and similarly for the other decay modes $n
  \to x^0 M$. These subsume the case where $x^0 = \bar\nu$ or $x^0=\nu$.} In
the right-most column of Table \ref{neutron_decay_table}, we list our estimates
for the lower bounds on $n \to \bar\nu \gamma$ obtained by combining the
relation (\ref{tau_n_to_nubar_gamma_wrt_nubar_meson}) with the experimental
lower bounds given in the middle column.  As is again evident from this table,
our approximate lower bounds on $(\tau/B)_{n \to \bar\nu \gamma}$ using the
experimental limits on $n \to \bar\nu \pi^0$ and $n \to \bar\nu \eta$ are
stronger than the bound from a direct experimental search, which is
$(\tau/B)_{n \to \bar\nu \gamma} > 0.550 \times 10^{33}$ yr \cite{takhistov15}.
%
%
%
%


\section{$p \to \ell^+ \ell'^+ \ell'^-$ Decays}
\label{p_to_3l_section}

\begin{figure*}
  \begin{center}
    \subfloat[]{
       \includegraphics[width=0.3\textwidth]{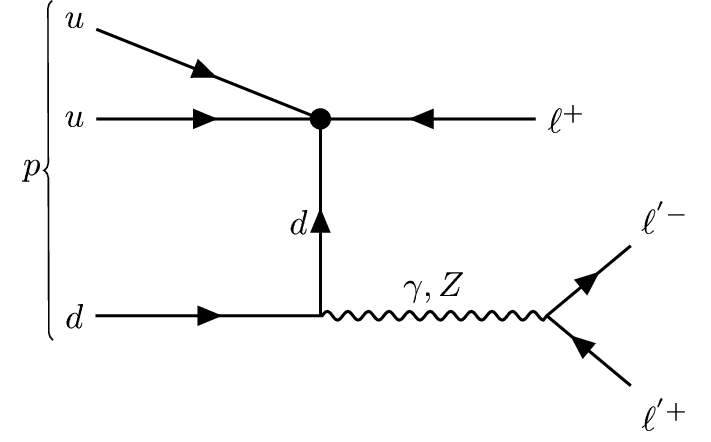}
     }
   \subfloat[]{
       \includegraphics[width=0.3\textwidth]{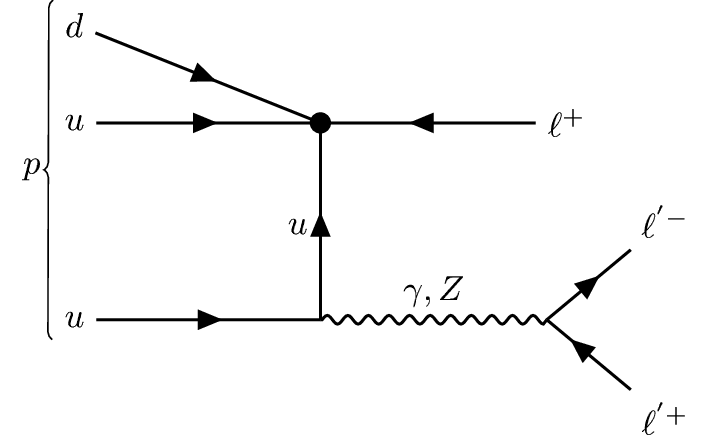}
     }
   \subfloat[]{
       \includegraphics[width=0.3\textwidth]{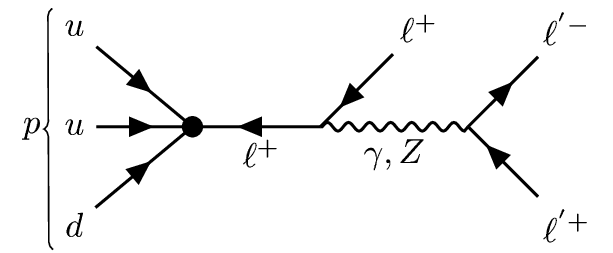}
     }
  \end{center}
\caption{Feynman diagrams for $p \to \ell^+ \ell'^+ \ell'^-$ with
$\ell, \ \ell' =e, \ \mu$.}
\label{p_to_lll_figure}
\end{figure*}

In this section we calculate estimated lower bounds on the partial lifetimes of
several proton decays of the form $p \to \ell^+ \ell'^+\ell'^-$, where $\ell$
and $\ell'$ denote $e$ or $\mu$, including both of the cases $\ell = \ell'$ and
$\ell \ne \ell'$. Graphs for the above-mentioned decays are shown in Fig.
\ref{p_to_lll_figure}. As discussed above, our theoretical framework for this
and our other estimates is a minimal one in which we assume only the
baryon-number-violating physics beyond the Standard Model that gives rise to
${\cal L}_{eff}$. (If one were to assume other BSM physics involving new
particles with masses much smaller than the GUT scale, then other graphs would
become relevant (e.g., \cite{hh})
%
%
%
%
We denote the four-momenta of the $p$, $\ell^+$, $\ell'^-$, and $\ell'^+$ as
$p$, $p_3$, $p_2$, and $p_1$, respectively, and set $q = p_1+p_2 = p-p_3$. Let
us first consider the $q \bar q$ annihilation producing a virtual photon, as
indicated in Figs. \ref{p_to_lll_figure}(a) and \ref{p_to_lll_figure}(b).  If
the total angular momentum of the $q\bar q$ system is $J= 0$, then the matrix
element for this sub-process is proportional to $q^\lambda [\bar
u(p_2)\gamma_\lambda v(p_1)]$ (where $\bar u(p_2)$ and $v(p_1)$ are Dirac
spinors), which vanishes. In the corresponding terms involving $Z$ exchange, if
the angular momentum $J$ of the $q \bar q$ subsystem is zero, then the matrix
element for the sub-process is proportional to $q^\lambda [\bar
u(p_2)\gamma_\lambda\{-(1/2)P_L+\sin^2\theta_W \}v(p_1)]$, where $P_L =
(1-\gamma_5)/2$.  In this case, the vector part of the neutral current gives
zero contribution and the axial-vector part gives a small term $\propto
2m_{\ell'}[\bar u(p_2)\gamma_5 v(p_1)]$, Thus, the dominant contributions from
these graphs are expected to arise from $q \bar q$ configurations with $J=1$.
For these $J=1$ terms, the contributions to the amplitude from the graphs with
a virtual photon are expected to dominate over the contributions from the
graphs with a virtual $Z$ by a factor $ \sim e^2/(G_F m_p^2) \sim 10^3$.  The
contribution from the virtual photon in the graph of
Fig. \ref{p_to_lll_figure}(c) is similarly dominant over that from the virtual
$Z$. It is thus natural to use, for comparison, proton decays to vector mesons,
$p \to e^+\rho^0$ and $p \to e^+ \omega$.  In contrast to the case with the $p
\to \ell^+ \gamma$ decays, this comparison connects a decay to a two-body final
state to a decay with a three-body final state. Because the integral
(\ref{decay_rate}) for the decay into a three-body final state involves details
of ${\cal L}_{eff}$ in a nontrivial integration, while the corresponding
integral for the decay into a two-body final state only involves a trivial
angular integration (since the magnitudes of the 3-momenta of the two
final-state particles are fixed), it is difficult to make a precise comparison
between the rates for these two decays. For a rough approximation, we will
simply take into account the differences in phase space for these decays, via
the ratios of the two-body and three-body phase-space factors.  Since the
three-body phase space factor has dimensions of (mass)$^2$ and the mass scale
is set by the initial nucleon mass, we will introduce a dimensionless
three-body phase space factor for a decay to a given final state $f.s.$,
denoted $\bar R^{(f.s.)}_3$, as defined in Eq. (\ref{rnbar}) in Appendix
\ref{phase_space_appendix}.  This quantity has the value $1/(2^8 \pi^3)$ if all
of the three final-state particles have zero or negligibly small
masses. Expressions for $(2^8 \pi^3)\bar R^{(f.s.)}_3$ for relevant final
states with non-negligible masses are given in
Eqs. (\ref{r3bar_m1m20})-(\ref{r3bar_mm0}) in Appendix
\ref{phase_space_appendix}.

We are thus led to the estimate
\beq
\Gamma_{p \to \ell^+ \ell'^+ \ell'^-} \sim (4\pi \alpha_{em})^2 \, 
\bigg [ \frac{\bar R_3^{(\ell^+ \ell'^+ \ell'^-)}}
             {     R_2^{(\ell^+ M)}} \bigg ] \, \Gamma_{p \to \ell^+ M} 
\label{rate_ptolll_relation}
\eeq
or equivalently, 
\beq
(\tau/B)_{p \to \ell^+ \ell'^+ \ell'^-} \sim (4\pi \alpha_{em})^{-2} \, 
\bigg [ \frac{     R_2^{(\ell^+ M)}             }
             {\bar R_3^{(\ell^+ \ell'^+ \ell'^-)} } \bigg ] 
\, (\tau/B)_{p \to \ell^+ M} 
\label{tau_ptolll_relation}
\eeq
Substituting the experimental lower bounds 
$(\tau/B)_{p \to e^+ \rho^0} > 0.720 \times 10^{33}$ yr. and
$(\tau/B)_{p \to e^+ \omega} > 1.60 \times 10^{33}$ yr. \cite{abe17d} 
in Eq. (\ref{tau_ptolll_relation}) and taking account of the ratios
of phase-space factors, we obtain estimated lower bounds on 
$(\tau/B)_{p \to e^+ e^+ e^-}$ of $0.9 \times 10^{37}$ yr. and 
$2 \times 10^{37}$ yr, respectively. These are 
much stronger than the lower bound on the partial lifetime for this 
decay from a direct experimental search, which is \cite{mcgrew99}
$(\tau/B)_{p \to e^+ e^+ e^-} > 0.793 \times 10^{33}$.  
Substituting the experimental lower bounds 
$(\tau/B)_{p \to \mu^+ \rho^0} > 0.570 \times 10^{33}$ yr. and
$(\tau/B)_{p \to \mu^+ \omega} > 2.80 \times 10^{33}$ yr. \cite{abe17d} 
in Eq. (\ref{tau_ptolll_relation}) and computing the ratios
of phase-space factors, we obtain estimated lower bounds on 
$(\tau/B)_{p \to \mu^+ e^+ e^-}$ of 
$0.6 \times 10^{37}$ yr and $3 \times 10^{37}$ yr, respectively. We
conservatively list these as $(\tau/B)_{p \to \mu^+ e^+ e^-} \gsim 10^{37}$ yr
in Table \ref{nucleon_decay_table}.  Again, these are much stronger 
than the experimental lower bounds from direct searches, namely 
$(\tau/B)_{p \to \mu^+ e^+ e^-} > 0.529 \times 10^{33}$ and
$(\tau/B)_{p \to \mu^+ \mu^+ \mu^-} > 0.675 \times 10^{33}$  \cite{mcgrew99}.
These estimated lower bounds on partial lifetimes are summarized in 
Table \ref{nucleon_decay_table}, in comparison with the current lower bounds
from direct experimental searches.  

With regard to these and other nucleon decay modes for which our estimates
yield lower bounds on the partial lifetimes that are much greater than existing
bounds from direct experimental searches, we stress that this does not mean
that these decay modes are not worth searching for in further experiments. If,
for example, in the future, the decays $p \to e^+ \pi^0$ and 
$p \to e^+ e^+ e^-$ are both observed and the value of 
$(\tau/B)_{p \to e^+ e^+ e^-}$ is significantly lower than a range 
estimated from Eq.
 (\ref{tau_ptolll_relation}), this would be doubly interesting, 
as evidence not only of baryon number violation incorporated in ${\cal
  L}_{eff}$, but also of additional relevant physics beyond the Standard 
Model. This comment also applies for the other nucleon decay
channels to be discussed below, for which our lower bounds on partial lifetimes
are much higher than the lower bounds from direct experimental searches. 


\section{ $n \to \bar\nu \ell^+ \ell^-$ and
$n \to \bar\nu \ell^+ \ell'^-$ Decays}
\label{n_to_nu2l_section}

\begin{figure*}
  \begin{center}
 \subfloat[]{
       \includegraphics[width=0.3\textwidth]{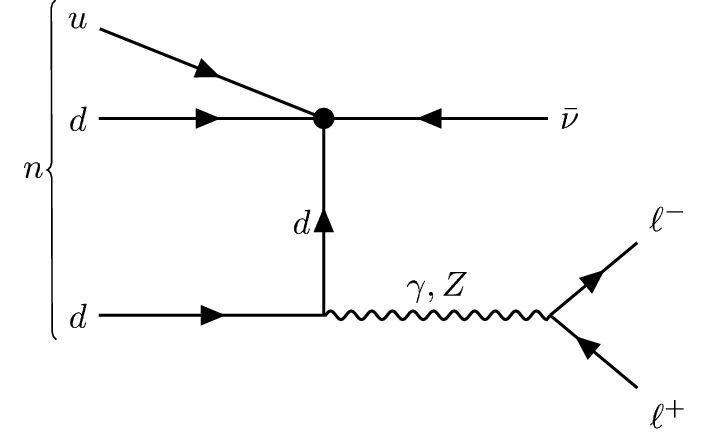}
     }
      \subfloat[]{
       \includegraphics[width=0.3\textwidth]{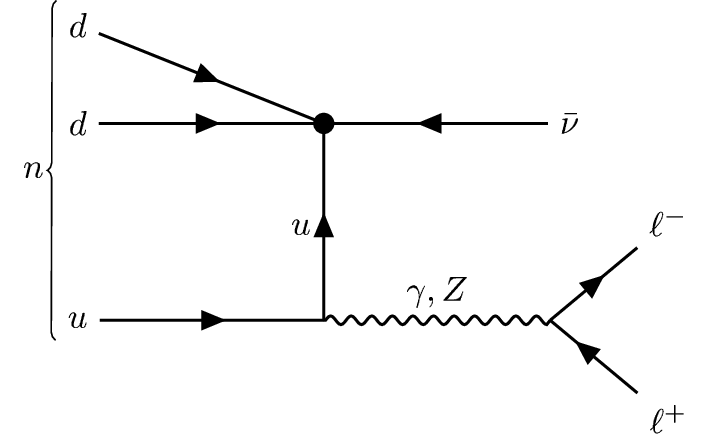}
     }
      \subfloat[]{
       \includegraphics[width=0.3\textwidth]{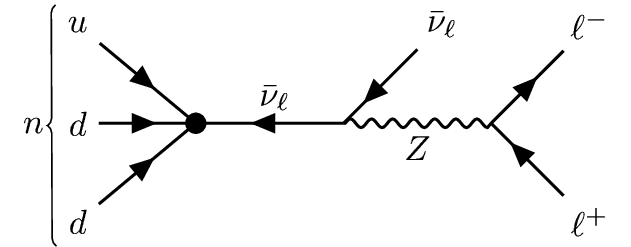}
     }
     \\
     \subfloat[]{
       \includegraphics[width=0.3\textwidth]{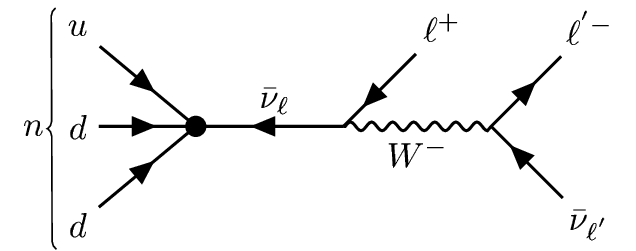}
     }
  \end{center}
\caption{Feynman diagrams for $n \to \bar\nu \ell^+ \ell^-$ and
 $n \to \bar\nu \ell^+ \ell'^-$ with $\ell, \ \ell' = e, \ \mu$.}
\label{n_to_nubar_ll_figure}
\end{figure*}

In this section we analyze the neutron decays $n \to \bar\nu \ell^+ \ell^-$ and
$n \to \bar\nu \ell^+ \ell'^-$, where $\ell, \ \ell' =e, \ \mu$. Feynman
diagrams for $n \to \bar\nu \ell^+ \ell^-$ decays are shown in Fig. 
\ref{n_to_nubar_ll_figure}.  In the graphs in Figs. 
\ref{n_to_nubar_ll_figure}(a) and \ref{n_to_nubar_ll_figure}(b), the $\bar\nu$
could be an EW-doublet or an EW-singlet antineutrino, while
in the graphs of Figs. \ref{n_to_nubar_ll_figure}(c) and
\ref{n_to_nubar_ll_figure}(d), the $\bar\nu = \bar\nu_\ell$ is an EW-doublet
antineutrino.  Since an experiment would not observe the $\bar\nu$, it would 
not distinguish between these possibilities. In the graphs of Figs.
\ref{n_to_nubar_ll_figure}(a)-\ref{n_to_nubar_ll_figure}(c), the charged
(anti)leptons are of the same generation, while in Fig.
\ref{n_to_nubar_ll_figure}(d), $\ell'^-$ may be of a different generation than
$\ell^+$.
%
%
%
%
As discussed before, the contributions of the diagrams with a virtual $Z$ in
Figs.  \ref{n_to_nubar_ll_figure}(a)-\ref{n_to_nubar_ll_figure}(c) are very
small compared with the contributions of the diagrams of Figs.
\ref{n_to_nubar_ll_figure}(a) and \ref{n_to_nubar_ll_figure}(b) with a virtual
photon.  This is also true of the graph with a virtual $W$ in
Fig. \ref{n_to_nubar_ll_figure}(d). Therefore, by arguments similar to those
used for the analysis of $p \to \ell^+ \ell'^+ \ell'^-$ decays, we estimate
\beq
(\tau/B)_{n \to \bar\nu \ell^+ \ell^-} \sim (4\pi\alpha_{em})^{-2} \, 
\bigg [ \frac{R_2^{(\bar\nu M)} }
            {\bar R_3^{(\bar\nu \ell^+ \ell'^-)} } \bigg ] \, 
(\tau/B)_{n \to \bar\nu M} \ . 
\label{n_to_nu_ll_wrt_n_to_num}
\eeq
For the same reasons discussed in connection with the $p \to \ell^+ \ell'^+
\ell'^-$ decays, when applying the relation (\ref{n_to_nu_ll_wrt_n_to_num}), we
will use $n \to \bar\nu M$ decays with $M$ being a vector meson, $\rho^0$ or
$\omega$.  As an illustrative example, we consider the decay $n \to \bar\nu e^+
e^-$.  Since $m_e^2/m_n^2 = 3.0 \times 10^{-7}$ is negligibly small, it follows
that, to very good accuracy, $\bar R_3 = \bar R_{3,0}$ (see
Eq. (\ref{r3_massless}).  Therefore, Eq. (\ref{n_to_nu_ll_wrt_n_to_num}) takes
the explicit form
\beqs
&& (\tau/B)_{n \to \bar\nu e^+e^-} \sim (4\pi\alpha_{em})^{-2} \, 
\bigg [ \frac{R_2^{(\bar\nu M)} }
            {\bar R_3^{(\bar\nu e^+e^-)} } \bigg ] \, 
(\tau/B)_{n \to \bar\nu M} \cr\cr
         &\sim& (4\pi\alpha_{em})^{-2} \, \bigg [ (2^5\pi^2)
\Big ( 1 - \frac{m_{M}^2}{m_n^2} \Big ) \bigg ] \, 
(\tau/B)_{n \to \bar\nu M} \ . 
\label{n_to_nu_ee_wrt_n_to_num}
\eeqs
Of the two $n \to \bar \nu M$ decay channels, the experimental lower bound 
on the channel with $M=\omega$ is the stronger one, so we focus on it. 
Substituting the lower bound $(\tau/B)_{n \to \bar\nu \omega} > 1.08 \times
10^{32}$ from Ref. \cite{mcgrew99} in Eq. (\ref{n_to_nu_ll_wrt_n_to_num}) and
evaluating Eq. (\ref{n_to_nu_ee_wrt_n_to_num}), we obtain the estimated lower
bound $(\tau/B)_{n \to \bar\nu e^+e^-} \gsim 1.2 \times 10^{36}$ yr. This is
much stronger than the lower bound from a direct search, namely $(\tau/B)_{n
  \to \bar\nu e^+e^-} > 2.57 \times 10^{32}$ yr.

Next, again using the same experimental lower bound on 
$(\tau/B)_{n \to \bar\nu \omega}$ and computing the ratio of phase-space
factors using Eqs. (\ref{r2}) and (\ref{r3bar_mm0}) in Appendix 
\ref{phase_space_appendix}, we obtain the estimated lower bound 
$(\tau/B)_{n \to \bar\nu \mu^+\mu^-} \gsim 1.6 \times 10^{36}$ yr.  This is a 
much more stringent lower bound than the one from a direct experimental search,
which is $(\tau/B)_{n \to \bar\nu \mu^+\mu^-} > 0.79 \times 10^{32}$ yr. 

Finally, we discuss the decay $n \to \bar\nu e^\pm \mu^\mp$.  Only the Feynman
diagram of Fig. \ref{n_to_nubar_ll_figure}(d) contributes to this decay, so we
obtain the estimate
\beq
(\tau/B)_{n \to \bar\nu e^\pm \mu^\mp} \sim (G_F m_n^2)^{-2} \, 
\bigg [ \frac{R_2^{(\bar\nu M)} }
            {\bar R_3^{(\bar\nu e^\pm \mu^\mp)} } \bigg ] \, 
(\tau/B)_{n \to \bar\nu M} \ . 
\label{n_to_nu_emu_wrt_n_to_num}
\eeq
Using the experimental lower bound on $(\tau/B)_{n \to \bar\nu \omega}$ again,
we obtain the estimate $(\tau/B)_{n \to \bar\nu e^\pm \mu^\mp} \gsim 1 \times
10^{44}$ yr.  This is much stronger than the bound from the direct search for a
decay of this type, namely $(\tau/B)_{n \to \bar\nu\mu^+e^-} > 0.83 \times
10^{32}$ yr. \cite{mcgrew99}.


\section{$p \to \ell^+ \nu \bar\nu$} 
\label{p_to_l2nu_section}

\begin{figure*}
  \begin{center}
      \subfloat[]{
       \includegraphics[width=0.3\textwidth]{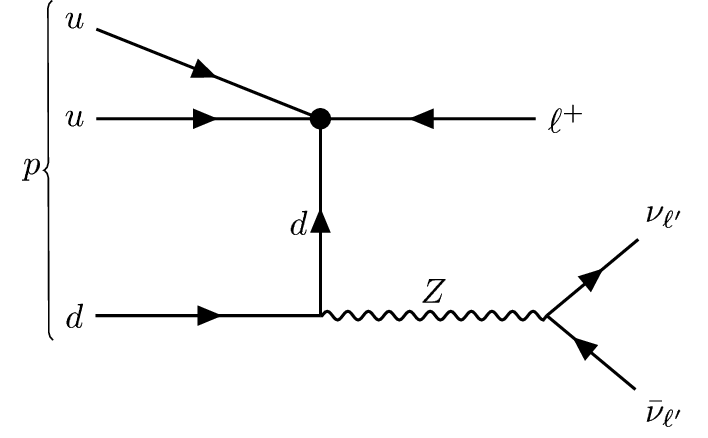}
     }
     \subfloat[]{
       \includegraphics[width=0.3\textwidth]{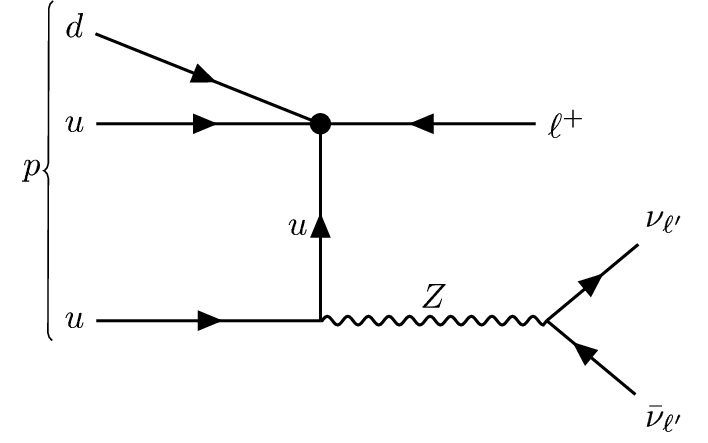}
     }
      \subfloat[]{
       \includegraphics[width=0.3\textwidth]{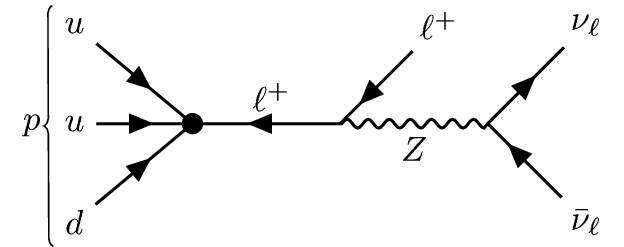}
     }
     \\
      \subfloat[]{
       \includegraphics[width=0.3\textwidth]{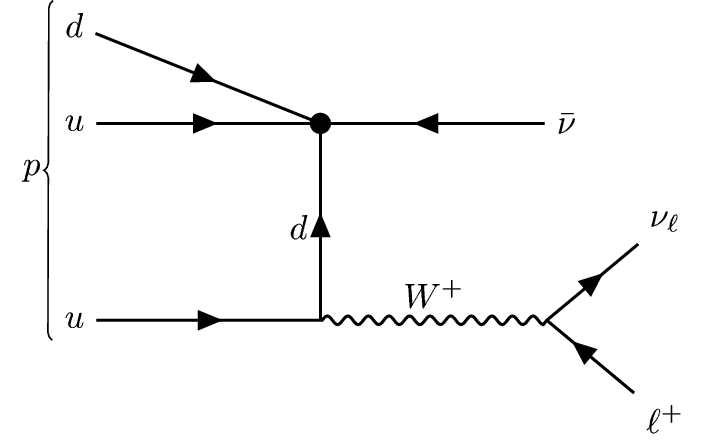}
     }
     \hspace{1 cm}
      \subfloat[]{
       \includegraphics[width=0.3\textwidth]{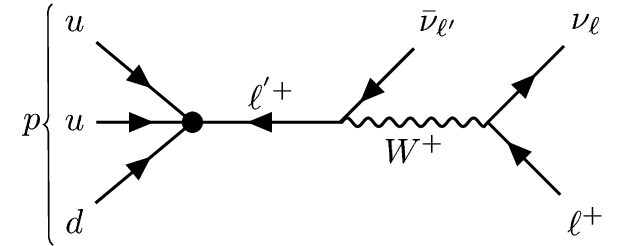}
     }
  \end{center}
\caption{Feynman diagrams for $p \to \ell^+ \nu \bar\nu$ with
$\ell^+=e^+, \ \mu^+$.}
\label{ptol2nu_figure}
\end{figure*}

In this section we derive an estimated bound for several different types of 
proton decays which are experimentally indistinguishable, namely 
(i)   $p \to \ell^+ \nu_{\ell'}\bar\nu_{\ell'}$;
(ii)  $p \to \ell^+ \nu_\ell \bar\nu_{\ell'}$; and
(iii) $p \to \ell^+ \nu_\ell \bar\nu_s$. We will refer collectively to these as
$p \to \ell^+ \nu \bar\nu$. Experimentally, these are all of the
form $p \to \ell^+ + {\rm missing}$, where ``missing'' denotes two
neutral weakly interacting particles, antiparticles, or a particle-antiparticle
pair, which do not decay in the detector. Experimental papers often use the
symbolic notation $p \to \ell^+ \nu\nu$ for all of these
decays.  Graphs that contribute to the decays (i)-(iii) are shown in Fig. 
\ref{ptol2nu_figure}. 
%
%
The graphs in Figs. \ref{ptol2nu_figure}(a)-\ref{ptol2nu_figure}(c) contain an
internal $Z$ line and contribute to decays of type (i) and the subset of the
decays of type (ii) in which $\nu_\ell = \nu_{\ell'}$.  The graphs in
Figs. \ref{ptol2nu_figure}(d) and \ref{ptol2nu_figure}(e) contain an internal
$W$ line and contribute to decays of the form (ii) and (iii), depending on
whether the $\bar\nu$ emitted from the effective four-fermion BNV vertex is an
EW-doublet antineutrino or an EW-singlet antineutrino.

Within the context of our theoretical framework, our relation for decays of
this type is
\beq
(\tau/B)_{p \to \ell^+ \nu \bar\nu} \sim (G_F m_p^2)^{-2} \, 
\bigg [ \frac{R_2^{(\ell^+ M)} }
            {\bar R_3^{(\ell^+ \nu \bar\nu)} } \bigg ] \, 
(\tau/B)_{p \to \ell^+ M} \ . 
\label{p_to_l2nu_wrt_p_to_lmeson}
\eeq
Using the experimental lower bounds on $(\tau/B)_{p \to \ell^+ \omega}$ given
in Table \ref{proton_decay_table}, we obtain the estimate $(\tau/B)_{p \to
  \ell^+ \nu \bar\nu} \gsim 10^{45}$ yrs for $\ell^+ = e^+, \ \mu^+$.  This is
much stronger than the lower bounds from direct searches, which are
$(\tau/B)_{p \to e^+ \nu \bar\nu} > 1.7 \times 10^{32}$ yr and $(\tau/B)_{p \to
  \mu^+ \nu \bar\nu} > 2.2 \times 10^{32}$ yr \cite{takhistov14}.


\section{$n \to \bar\nu \bar\nu \nu$} 
\label{n_to_3nu_section}

\begin{figure*}
  \begin{center}
      \subfloat[]{
       \includegraphics[width=0.3\textwidth]{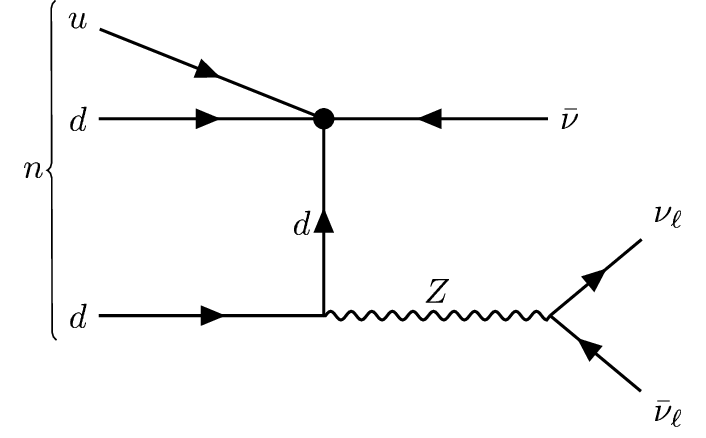}
     }
     \subfloat[]{
       \includegraphics[width=0.3\textwidth]{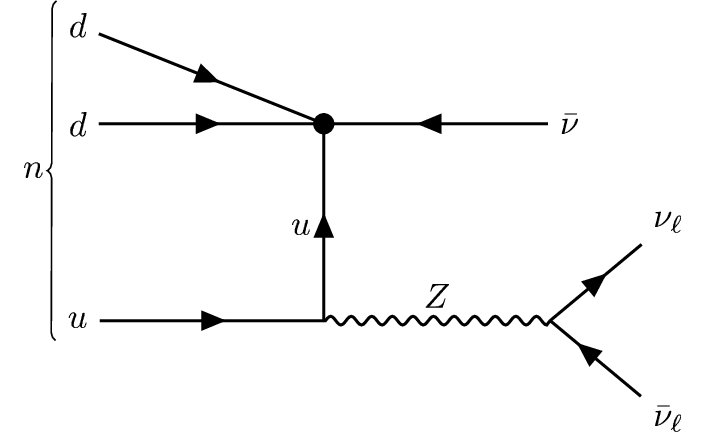}
     }
      \subfloat[]{
       \includegraphics[width=0.3\textwidth]{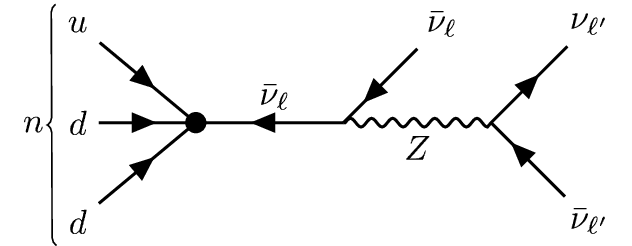}
     }
  \end{center}
\caption{Feynman diagrams for $n \to \bar\nu \bar\nu \nu$ decay.}
\label{nto3nu_figure}
\end{figure*}

Finally, we consider the decays of the generic form 
$n \to \bar\nu \bar\nu\nu$. Graphs that contribute to these decays are 
shown in Fig. \ref{nto3nu_figure}. 
%
%
In the processes depicted in Figs. \ref{nto3nu_figure}(a) and
\ref{nto3nu_figure}(b), the $\bar\nu$ that emanates from the BNV four-fermion
vertex can be either an EW-doublet antineutrino (of any flavor) or an
EW-singlet antineutrino, while the $\nu\bar\nu$ pair produced by the virtual
$Z$ are EW-doublet (anti)neutrinos.  In Fig. \ref{nto3nu_figure}(c), the two
antineutrinos and the neutrino are all of EW-doublet type.  We estimate
\begin{widetext}
\beq
(\tau/B)_{n \to \bar\nu \bar\nu \nu} \sim (G_F m_n^2)^{-2} \, 
\bigg [ (2^5\pi^2) \Big ( 1 - \frac{m_{M}^2}{m_n^2} \Big ) \bigg ] \, 
(\tau/B)_{n \to \bar\nu M} \ . 
\label{n_to_3nu_wrt_n_to_nu_meson}
\eeq
\end{widetext}
Using the experimental lower limit
$(\tau/B)_{n \to \bar\nu \omega} > 1.08 \times 10^{32}$ yr \cite{mcgrew99}, as
listed in Table \ref{proton_decay_table}, we obtain the estimate $(\tau/B)_{n
  \to \bar\nu\nu\bar\nu} \gsim 10^{44}$ yr. This is much stronger than the
lower bound from a direct experimental search, namely 
$(\tau/B)_{n \to {\rm inv.}} > 0.58 \times 10^{30}$ yr \cite{araki06}. This
experimental lower bound was set by the KamLAND experiment from a search for 
gamma rays from the de-excitation of the ${}^{11}$C nucleus that would 
result from the $n \to {\rm inv.}$ decay of a neutron in a 
${}^{12}$C atom in its liquid scintillator detector.  


\section{Other Decay Modes}
\label{other_decays_section} 

Our method can also be applied to other nucleon decay modes, such as 
$p \to \ell^+ \gamma\gamma$ and $n \to \bar\nu \gamma\gamma$, where here the
$\gamma\gamma$ part of the final state are ``continuum'' photons, i.e., not
photons that arise from a cascade decay such as $p \to \ell^+ \pi^0$ followed
by $\pi^0 \to \gamma\gamma$ or $p \to \ell^+ \eta$ followed by $\eta \to
\gamma\gamma$.  Graphs for these decays with continuum diphotons are obtained
from those for $p \to \ell^+ \gamma$ in Fig. \ref{p_to_l_gamma_figure} and
for $n \to \bar\nu \gamma$ in Fig. \ref{n_to_nubar_gamma_figure} and hence are
not shown separately. Using the same methods as before, we obtain 
\beq
(\tau/B)_{p \to \ell^+ \gamma\gamma} \sim (4\pi \alpha_{em})^{-2} \, 
\bigg [ \frac{R^{(\ell^+ M)}_2}{\bar R^{(\ell^+ \gamma\gamma)}_3} \bigg ] 
\, (\tau/B)_{p \to \ell^+ M}
\label{tau_p_to_l2gamma_wrt_lmeson}
\eeq
and
\beqs
(\tau/B)_{n \to \bar\nu^+  \gamma\gamma} &\sim& 
(4\pi \alpha_{em})^{-2} \, 
\bigg [ \frac{R^{(\bar\nu M)}_2}{\bar R^{(\bar\nu \gamma\gamma)}_3} \bigg ] 
\, (\tau/B)_{n \to \bar\nu M} \cr\cr
&\sim& 
\bigg [ (2^5\pi^2) \Big ( 1 - \frac{m_{M}^2}{m_n^2} \Big ) \bigg ] 
\,  (\tau/B)_{n \to \bar\nu M} \ . \cr\cr
&& 
\label{tau_n_to_nu2gamma_wrt_lmeson}
\eeqs
Using the experimental lower bounds on $p \to e^+ M$ and $p \to \mu^+ M$
listed for $M = \rho^0, \ \omega$ in Table \ref{proton_decay_table}, the more
stringent of which are for $p \to \ell^+ \omega$, we obtain the estimates
$(\tau/B)_{p \to e^+ \gamma\gamma} \gsim 2 \times 10^{37}$ yr.  and
$(\tau/B)_{p \to \mu^+ \gamma\gamma} \gsim 3 \times 10^{37}$ yr.  An
experimental lower bound from a direct search is $p \to
e^+ \gamma\gamma$ decay mode, namely $(\tau/B)_{p \to e^+ \gamma\gamma} >
1.0 \times 10^{32}$ yr. \cite{berger91}. Our estimated lower bound is much
stronger than this direct limit.  We are not aware of any 
published experimental lower bound on $(\tau/B)_{p \to \mu^+ \gamma\gamma}$. 

By similar methods, we obtain the estimated lower bound $(\tau/B)_{n \to
  \bar\nu \gamma\gamma} \gsim 10^{36}$ yr. An experimental lower bound is
$(\tau/B)_{n \to \bar\nu \gamma\gamma} > 2.39 \times 10^{32}$ yr from the
IMB3 experiment \cite{mcgrew99}.  This was an inclusive search for any events
of this type, which also allowed for the possibility that the invariant
diphoton mass was equal to $m_{\pi^0}$ or $m_\eta$ to within the detector
resolution \cite{mcgrew_pv}. Our estimated lower bound is again much stronger
than this direct limit.  These results are summarized in Table
\ref{nucleon_decay_table}.  One can also apply these techniques to relate other
baryon-violating processes to each other. We have done this to derive improved
bounds on certain $\Delta B=-2$ dinucleon decays.  These results are reported
elsewhere.


\section{Conclusions}
\label{conclusion_section} 

In this paper we have calculated estimated lower bounds on the partial
lifetimes for several nucleon decays, including $p \to \ell^+ \ell'^+ \ell'^-$,
$n \to \bar\nu \ell^+ \ell'^-$, $p \to \ell^+ \nu \bar\nu$, and $n \to \bar\nu
\bar\nu \nu$, where $\ell$ and $\ell'$ denote $e$ or $\mu$. We assume a minimal
theoretical framework in which the only physics beyond the SM is that which
produces the four-fermion operators in the baryon-number-violating effective
Lagrangian responsible for these nucleon decays. Our method relies on relating
the rates for these decay modes to the rates for decay modes of the form $p \to
\ell^+ M$ and $n \to \bar\nu M$, where $M$ is a pseudoscalar or vector meson,
and then using the experimental lower bounds on these latter decays.  Although
our estimates are rough, our lower bounds are substantially stronger
than lower bounds on the partial lifetimes for these decays from direct
experimental searches.  We also
present corresponding estimated
 lower bounds on partial lifetimes for the radiative
decays $p \to \ell^+\gamma$, $n \to \bar\nu \gamma$, $p \to \ell^+
\gamma\gamma$, and $n \to \bar\nu \gamma\gamma$.  
There are strong motivations for pushing
the search for nucleon decay to greater sensitivity in many channels.  It is
hoped that this search will be carried out with current data and with 
future nucleon decay experiments. 


\begin{acknowledgments}

This research was supported in part by the NSF Grants NSF-PHY-1620628 and 
NSF-PHY-1915093 (R.S.). 

\end{acknowledgments}


\begin{appendix}

\section{Phase Space Formulas} 
\label{phase_space_appendix}

In general, the decay rate of a parent particle $N$ with 
four-momentum $p$ satisfying $p^2=m_N^2$ to a final state ($f.s.$) consisting 
of $n$ particles with four-momenta $p_i$, $1 \le i \le n$, is given by 
\beq
\Gamma_{N \to f.s.} = 
\frac{S}{2m_N} \, \int dR_n \, |A_{N \to f.s.}|^2 \ , 
\label{decay_rate}
\eeq
where a sum over polarizations of final-state particles and an average over the
polarizations of the parent particle are understood; 
$A_{N \to f.s.}$ denotes the amplitude for the decay;
$S$ is a symmetry factor to take account of possible identical particles in 
the final state; and the integration over the $n$-body final-state phase space
is given by 
\beq
\int dR_n = \frac{1}{(2\pi)^{3n-4}} \,
\int \Big [ \prod_{i=1}^n 
\frac{d^3 p_i}{2E_i} \Big ] \, \delta^4 \Big ( p-(\sum_{i=1}^n p_i) \Big ) \ .
\label{phase_space_integral}
\eeq
For the nucleon decays of interest here we have $m_N$ with $N=p$ or $N=n$,
and we denote $p_i^2=m_i^2$.

It is useful to consider the phase space integration by
itself, defining an $n$-body phase space factor $R_n$ as
\beq
R_n = \int dR_n 
\label{nbody_phasespace}
\eeq
For our applications, we will sometimes want to explicitly indicate the
final-state particles, and for this purpose, we will use the notation 
$R_n^{(f.s.)}$.  For example, for $p \to e^+ \pi^0$, this phase space factor is
written as $R_2^{(e^+\pi^0)}$, and so forth for other decays. 

The quantity $R_n$ has mass dimension $2(n-2)$, so we define a dimensionless
phase space factor 
\beq
\bar R_n \equiv (m_N)^{-2(n-2)}R_n \ . 
\label{rnbar}
\eeq
In general, 
\beq
R_2 = \frac{1}{8\pi} \, [\lambda(1,\delta_1,\delta_2)]^{1/2} 
\label{r2}
\eeq
where
\beq
\lambda(x,y,z) = x^2+y^2+z^2-2(xy+yz+zx) 
\label{lam}
\eeq
and 
\beq
\delta_i = \Big (\frac{m_i}{m_N} \Big )^2 \ .
\label{deltai}
\eeq

In the case where $m_i^2/m_N^2 << 1$ for all $i$, we denote the resultant 
$R_n$ as $R_{n,0}$. A general formula is 
\beq
\bar R_{n,0} = 
\frac{1}{2^{4n-5} \pi^{2n-3} \Gamma(n) \Gamma(n-1)} \quad {\rm for} \ n \ge 2 \
, 
\label{rn}
\eeq
where $\Gamma(n)$ is the Euler gamma function. In particular, For $n=2$ and
$n=3$,
\beq
R_{2,0} = \bar R_{2,0} = \frac{1}{2^3 \pi} 
\label{r2_massless}
\eeq
and
\beq
\bar R_{3,0} = \frac{1}{2^8 \pi^3} \ . 
\label{r3_massless}
\eeq

For final states of nucleon decays in which a $e^\pm$ occurs, its mass
satisfies the above condition of being negligibly small with respect to
$m_N$. For leptonic final states involving one or two $\mu^\pm$ and a third
particle of zero or negligibly small mass, we will make
use of two formulas for $\bar R_3$.  These follow from the general formula for
the three-body phase space, $\bar R_3$ with one massless final-state particle,
which is \cite{byckling_kajantie}
\begin{widetext}
\beqs
(2^8 \pi^3) \bar R_3(m_1,m_2,0) &=& (1+\delta_1 +
\delta_2) \, [\lambda(1,\delta_1,\delta_2)]^{1/2} 
+ 2|\delta_1-\delta_2| \, \ln \bigg [ 
\frac{\delta_1+\delta_2-(\delta_1-\delta_2)^2 + 
|\delta_1-\delta_2| \, [\lambda(1,\delta_1,\delta_2)]^{1/2}}
{2\sqrt{\delta_1\delta_2}} \bigg ] \cr\cr
&-&2 (\delta_1+\delta_2-2\delta_1\delta_2) \, 
\ln \bigg [ \frac{1-\delta_1-\delta_2+[\lambda(1,\delta_1,\delta_2)]^{1/2}}
{2\sqrt{\delta_1 \delta_2}} \bigg ] \ . 
\label{r3bar_m1m20}
\eeqs
\end{widetext}
Note that this is symmetric under the interchange $m_1 \leftrightarrow m_2$ 
and thus $\delta_1 \leftrightarrow \delta_2$. 

The first special case of (\ref{r3bar_m1m20}) that we will need is for
$m_1=m$, with $m_2$ and $m_3$ zero or negligibly small, so 
$\delta_1=(m/m_N)^2 \equiv \delta$. This case
applies for decays such as $p \to \mu^+ e^+e^-$ and 
$n \to \bar\nu \mu^\pm e^\mp$.
For this case we have
\beq
(2^8 \pi^3) \bar R_3(m,0,0) = 1-\delta^2 -2\delta \, 
\ln \Big ( \frac{1}{\delta} \Big ) \ . 
\label{r3bar_m00}
\eeq
Numerically, the right-hand side of Eq. (\ref{r3bar_m00}) has the value
0.889 for $m=m_\mu$ with parent particle $p$ or $n$. 
The second special case of (\ref{r3bar_m1m20}) that we will need is for
$m_1=m_2=m$, so $\delta_1=\delta_2 = (m/m_N)^2 \equiv \delta$.  This case
applies for decays such as $p \to e^+ \mu^+\mu^-$, $n \to \bar\nu \mu^+\mu^-$.
We have
\beqs
(2^8 \pi^3) \bar R_3(m,m,0) &=& (1+2\delta) \, \sqrt{1-4\delta} \cr\cr
&-&4\delta(1-\delta) \, \ln \bigg [ \frac{1-2\delta + \sqrt{1-4\delta}}
{2\delta} \bigg ] \ . \cr\cr
&&
\label{r3bar_mm0}
\eeqs
Numerically, the right-hand side of Eq. (\ref{r3bar_mm0}) has the value
0.782 for $m=m_\mu$ with parent particle $p$ or $n$.  

Since the form of ${\cal L}_{eff}$ in Eq. (\ref{leff}) depends on the unknown
details of the baryon-number-violating BSM physics, a knowledge of this BSM
physics would be necessary to calculate the full convolution of $|A|^2$
weighted with the three-body phase space in the $n=3$ case of
Eq. (\ref{decay_rate}).  This may be contrasted with the case in $\mu$ decay to
massive neutrinos, where such calculations have been performed for both $V-A$
charged currents and general Lorentz structure \cite{shrock81}.

\end{appendix}



\end{document}